\documentclass{emulateapj}
\usepackage{natbib}
\usepackage{graphicx}
\usepackage{amsmath}
\usepackage{multirow}
\usepackage{enumitem}
\usepackage[letterpaper,top=4cm,bottom=-.5cm,right=.8cm,left=2.2cm]{geometry}

\shorttitle{Quiet Sun Transition Region}
\shortauthors{Schmit \& De Pontieu}

\begin{document}

\title{What Is the Source of Quiet Sun Transition Region Emission?}
\author{D.J. Schmit\altaffilmark{1,}\altaffilmark{2}}
\author{Bart De Pontieu\altaffilmark{1,}\altaffilmark{3}}
\affil{Lockheed-Martin Solar and Astrophysics Laboratory, Palo Alto, CA 94304, USA}
\affil{Bay Area Environmental Research Institute, Sonoma, CA 94952, USA}
\affil{Institute of Theoretical Astrophysics, University of Oslo, P.O. Box 1029 Blindern, N-0315 Oslo, Norway}
\begin{abstract}
Dating back to the first observations of the on-disk corona, there has been a qualitative link between the photosphere's magnetic network and enhanced transition-temperature plasma emission.
These observations led to the development of a general model that describes emission structures through the partitioning of the atmospheric volume with different magnetic loop geometries that exhibit different energetic equilibria.
Does the internetwork produce transition-temperature emission?
What fraction of network flux connects to the corona?
How does quiet sun emission compare with low-activity Sun-like stars?
In this work, we revisit the canonical model of the quiet sun, with high-resolution observations from IRIS and HMI in hand, to address those questions.
We use over 900 deep exposures of Si IV 1393\AA~from IRIS along with nearly simultaneous HMI magnetograms to quantify the correlation between transition-temperature emission structures and magnetic field concentrations through a number of novel statistics.
Our observational results are coupled with analysis of the Bifrost MHD model and a large-scale potential field model. 
Our results paint a complex portrait of the quiet sun.
We measure an emission signature in the distant internetwork that cannot be attributed to network contribution.
We find that the dimmest regions of emission are not linked to the local vertical magnetic field.
Using the MHD simulation, we categorize the emission contribution from cool mid-altitude loops and high-altitude coronal loops and discuss the potential emission contribution of spicules.
Our results provide new constraints on the coupled solar atmosphere so that we can build on our understanding of how dynamic thermal and magnetic structures generate the observed phenomena in the transition region.
\end{abstract}
\section{Introduction}
Magnetic fields in the solar atmosphere exist in very different equilibria depending on altitude and strength.
At the photosphere, the magnetic field is largely isolated into tight concentrations due to the relative values of magnetic pressure and gas pressure.
The corona, alternatively, is structured on all scales by the magnetic field, which is volume filling and dominates force equilibrium and energy balance (through anisotropic conduction).
Magnetic fields in the transition region and chromosphere occupy an intermediate state and are not well understood.\\
\indent The large-scale pattern of photospheric magnetic field forms a network of mixed polarity flux concentrations that largely coincide with supergranular lanes \citep{simon_64}.
Network magnetic elements underlie regions of enhanced transition region and coronal emission \citep{reeves_76}.
The canonical structural model of the magnetic field that extends from these concentration was first suggest by \cite{gabriel_76} and expanded by \cite{dowdy_86}.
The field emanating from network elements form large-scale funnels which connect to the high-altitude atmosphere.
Underneath these funnels, nearby network regions of opposite flux are connected by lower altitude loops.
Using differential emission measure analysis, Gabriel found there was too much observed transition region emission to be explained by emission exclusively within funnels.
One proposed solution to that problem is that short loops reach maximum temperatures between $5\times 10^4$ K and $1\times10^5$ K \citep{feldman_83,dowdy_86}.
These loops were dubbed unresolved fine structure (UFS).
These structural models consider time-independent magnetic features which are undoubtedly a simplification from the real Sun.
Another route to produce the missing emission measure are dynamics related to spatially and temporally dependent heating.
1D hydrodynamic models suggest that if the density in the transition region varies significantly over unresolved timescales, the density-squared dependence of line emission can produce the extra emission \citep{wikstol_98}.
Considering the prevalence of spicules, dynamic features that are most commonly observed at chromospheric temperatures but contain a transition temperature component \citep{rouppe_15}, time-dependent effects must surely play a role.\\
\indent For observational constraints on the transition region, the Solar Ultraviolet Measurements of Emitted Radiation (SUMER) instrument  has been the most sensitive spectrograph.
Deep exposures using the photon-counting detector have been used for important studies in \cite{feldman_99} and \cite{warren_00}.
Using large-scale raster scans, Feldman et al found that the contrast between internetwork and network (based on a two-component log-normal distribution) varies as a function of ion temperature, with a peak near $10^5$ K.
EUV broadband imagers, the Transition Region and Coronal Explorer (TRACE) and the Atmospheric Imaging Assembly (AIA), have shown that while plasma emission above 1 MK is diffuse and omnipresent in the quiet sun, transition-temperature plasma is clumpy and isolated to patches near network.
These instruments are primarily sensitive to plasma around $8\times10^5$ K.
Using Interface Region Imaging Spectrograph (IRIS, \citealt{depontieu_14}) data, \cite{hansteen_14} were the first authors to report direct observations of the UFS.
They found that the data showed short ($<3$ Mm in height and $<10$ Mm in length) and bright loops in the FUV slit jaw images.
Those authors were able to find several highly contrasted but short lived examples, but  it remains unclear how much emission measure the UFS loops contribute (in a statistical sense) and how photospheric magnetic fields are related.\\
\indent Although the literatures have been developed separately, there must be a connection between the spicules and loops.
The morphology of spicules is reminiscent of segments of high-altitude coronal loops; they are collimated and generally exhibit little curvature in Ca II (we do not observe closed loops with any regularity).
Spicules exist in two varieties: a shorter ($<$8 Mm in length), longer lived but infrequent Type 1 and a more dynamic Type 2 \citep{depontieu_07b}.
Type 2 spicules can generate emission up to 1 MK \citep{skogsrud_15,depontieu_11} and are rooted primarily in network magnetic fields \citep{pereira_14}.
Spicules are the manifestation of a heating and mass-loading cycle connecting the chromosphere and corona.
Any loop model which does not produce spicules cannot be accurately capturing the physics of the solar atmosphere.\\
\indent Models suggest that the peak emission of the Si IV lines at 1393.8\AA~and 1402.8\AA~should occur between 5$\times10^4$ K$<T<8\times10^4$ K \citep{avrett_13,doschek_97} assuming ionization equilibrium.
Hydrostatic loop models suggest that plasma within this temperature range must exist in one of two energetic equilibria: conductively heated from the hot corona or short isothermal loops with weak heating \citep{antiochos_86}.
These loops types are incorporated into the two source theory of transition temperature line emission described in the Dowdy model: small filling factor coronal funnels and low emission measure but ubiquitous loops.
Of course, hydrostatic models are an inherent simplification of the chromosphere-corona system.
Magnetohydrodynamics (MHD) models have been constructed to study the joint evolution of thermal properties and magnetic field in plasma.
Coronal loops in time-dependent MHD models have been shown to be highly dynamic \citep{bourdin_13}.
The Bifrost stellar atmosphere code \citep{gudiksen_11} can simulate a self-consistent and linked chromosphere/corona.
\cite{hansteen_14} found dynamic cool loops with lifetimes of approximately 1-2 minutes in a Bifrost simulation.
While a single example of a spicule was produced by the Bifrost model \citep{jms_13}, it is unknown why this ubiquitous phenomenon does not frequently occur in the model. 
In our time-independent analysis we will be unable to differentiate between dynamic and static features.
Our goals focus on measuring time-averaged quantities over wide fields of view that provide the necessary big-picture context and will allow us to properly interpret the individual roles of dynamic features like spicules and UFS in future work.\\
\indent Magnetic connectivity is of central importance for linking the Sun and the heliosphere.
Field lines that thread the upper atmosphere act as conduits for mass and energy.
If you want to accurately constrain what the heat source for the quiet sun corona is, you must know where it connects to.\\
\indent The IRIS dataset offer us a new perspective on transition temperature-plasma (TTP) emission.
The Si IV lines, 1393.8\AA~ and 1402.8\AA, are resonant transitions from an abundant element that are strong emitters in this temperature regime.
We use the 1393\AA~line to re-examine the structure of the quiet sun transition region.
We compare Si IV radiance structures with the location and strength of magnetic elements extracted from photospheric magnetograms.
Our observational data are compared with a MHD model and a potential field model to interpret how magnetic geometries are linked to emission.
In Section 2, we discuss the IRIS observations and how we collate the HMI magnetograms.
In Section 3, we discuss how we determine the minimum detectable line emission for the IRIS data and how we extract magnetic elements from the magnetograms.
In Section 4, we study the correlation of statistics of the photospheric magnetic field and Si IV radiance.
In Section 5, we analyze the distribution of magnetic loop geometries within two magnetic models of the quiet sun.
In Section 6, we discuss our interpretation on the magnetic structure of the quiet sun based on the data and the models.
\section{Observations}
Our analysis includes four IRIS datasets taken using different observing modes, the details of which are presented in Table 1.
The primary factor in choosing these datasets were the long exposures and the small raster steps (where the slit raster step equals the slit width).
Preference was given to large field of view observations which primarily cover quiet sun.
DS2 contains some plage in the lower 30" of the raster (these spectra are omitted from the statistical analysis).
The term ``quiet sun" is rather nebulous, given the power law distribution of both magnetic flux density and magnetic flux at the photosphere.
There is an ill-defined line between network and decayed plage to which DS4 is likely close.
DS1 and DS2 were collected with solar rotation tracking disabled, thus the effective step size is smaller than the slit width (at sun center the rotation rate is about 0.09" over 30 seconds).\\
\indent IRIS Level 2 data is used in this analysis; it has been reduced via the process described in \cite{depontieu_14}.
We use FUV slit jaw data for alignment purposes and the FUV spectra for analysis.
The IRIS FUV spectrograph data is known to have contribution from optical/infrared photons.
This component to the spectra is nearly uniform across the spectrograph CCD.
It is subtracted out of IRIS FUV spectra during calibration, but it is a source of noise.
\begin{table*}
\centering
\begin{tabular}{cc|ccccc|c}
Name& Date & Spec. Bin& X-Bin& Y-Bin& Exp. Time & Eff. Area & OBSID\\
\hline
DS1& 2014-08-21 & 25.4 & 0.332 & 0.166 & 30 & 0.97 & 3800262196\\
DS2& 2014-03-05 & 25.4 & 0.332 & 0.332 & 30 & 1.22 & 3830113696\\
DS3& 2015-07-10 & 25.4 & 0.332 & 0.332 & 60 & 1.04 & 3600114057\\
DS4& 2015-07-12 & 25.4 & 0.332 & 0.332 & 60 & 1.05 & 3600114056\\
\end{tabular}
\caption{Description of IRIS datasets. Spectral binning (units of m\AA), angular size of slit/X spatial abscissa (units of arcsec), angular size of Y spatial abscissa, exposure times (units of seconds), effective area of the telescope (units in cm$^2$), observing program.}
\label{tab:ds}
\end{table*}
We elaborate on our measurement technique and associated uncertainties in Section 3.\\
\indent In this analysis, we compare structures in the transition region with photospheric magnetograms.
The photospheric data is provided by the Helioseismic and Magnetic Imager instrument (HMI, \citealt{scherrer_12}) aboard the Solar Dynamics Observatory (SDO).
HMI produces a map of magnetic flux density (the component along the line of sight, $B_{los}$) over the full sun every 45 seconds based on the line width of the Fe I 6173\AA~Stokes V profile.
The IRIS data is aligned with HMI by first aligning with SDO/AIA 1600\AA~(which has 24 second cadence).
The 1600\AA~channel of AIA uses a broadband filter that while it is centered on a transition region line, C IV at 1548\AA, includes a significant amount of continuum emission. 
This also holds true for the IRIS FUV slit jaw data, which is visibly very similar in structure to AIA 1600\AA.
A cross correlation algorithm is used to find the location of IRIS SJI image within the AIA image.
These coordinates are used to create a synthetic magnetogram.
Interpolations occur over multiple steps in this process.
Most IRIS observing modes do not expose the FUV SJI at every raster position.
To fill in the gaps, a linear interpolation is applied using the PZT keywords (which refer to the piezoelectric tranducers that are responsible for fine pointing) in the data headers as abscissa.
The Level 1 plate scale for HMI is 0.6"/pixel so the magnetic dataset is oversampled to IRIS resolution.
All four datasets have been processed in this manner automatedly.
\section{Data Reduction}
\subsection{IRIS detection limit}
Our first step in this analysis is deriving the line radiance for the Si IV 1393\AA~line.
We use a Gaussian line profile plus a constant background to fit 1.04\AA~wide window centered on 1393.6\AA.
\begin{figure*}
\centering
\includegraphics[width=0.7\textwidth]{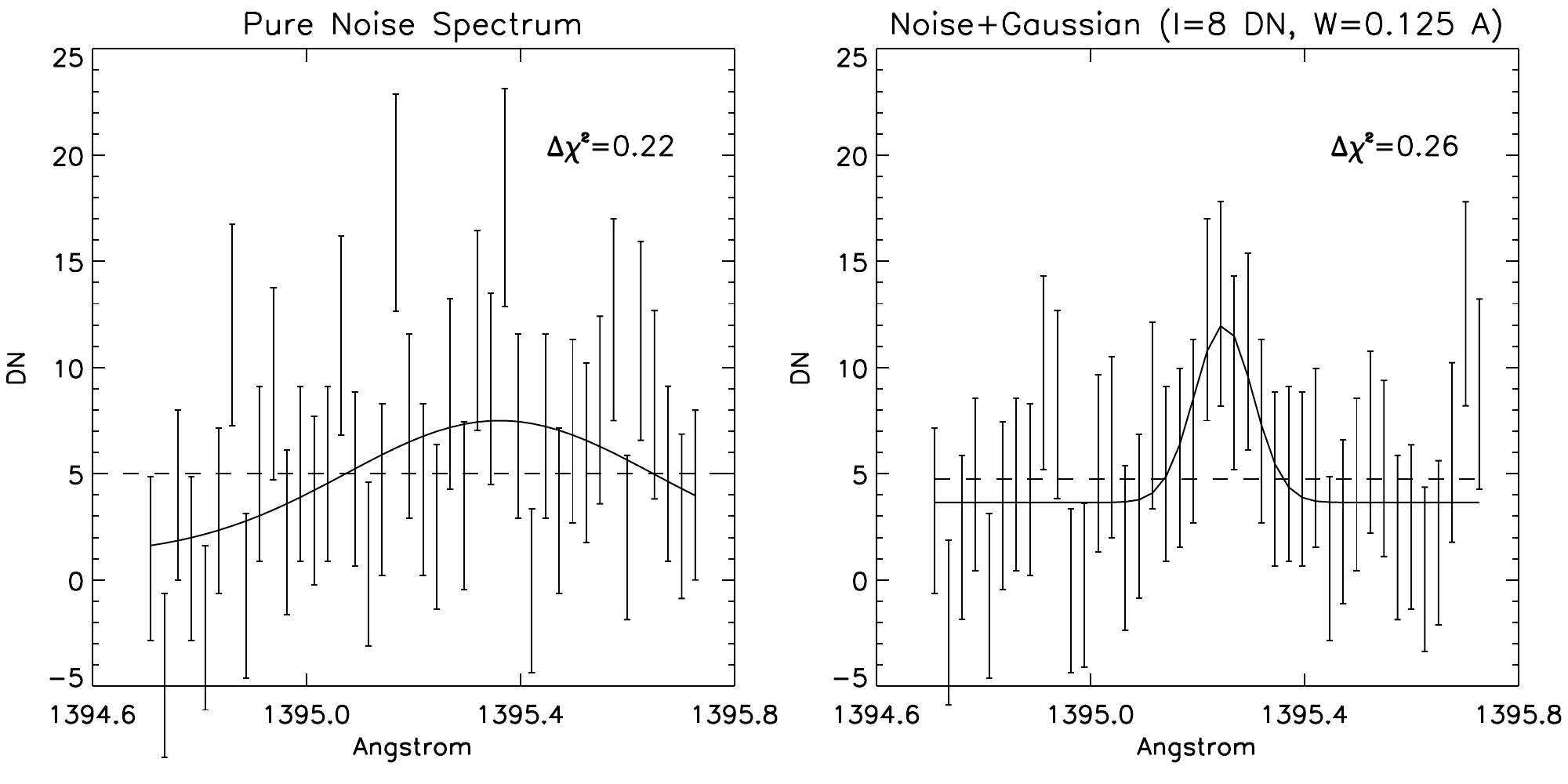}
\caption{Example spectra and best fit model for Monte Carlo detection limit calculation. (a) Pure noise and (b) noise plus a Poisson-convoluted gaussian line profile. }
\label{fig:noise}
\end{figure*}
We assume the errors for each spectral pixel are a linear combination of the Poisson noise (4 photons per DN) and 3 DN readout noise background \citep{depontieu_14}.
We attempt to remove cosmic ray and hot pixels by flagging pixels that exceed an intensity threshold relative to neighbors.
The MPFIT algorithm \citep{markwardt_09} is used to minimize $\chi^2$ and find a best fit model.
A robust error analysis must assess which of those models are accurate depictions of the data.
We are also interested in determining the minimum detection limit (i.e. what is the minimum number of photons required across the line that we can reject the null hypothesis that noise would lead to the given best fit model).
We have taken measures to quantify these uncertainties.
First, we measure the rate of false positives (i.e. how often does the algorithm return a model of given integrated intensity when the spectrum is pure noise) using a 1\AA~spectral band centered at 1395.2\AA~devoid of any previously measured lines .
Next, we measure the rate of true positives (i.e. how often will a line embedded in noise be properly measured) by embedding a Gaussian of known radiance and centroid but subject to Poisson noise into the example IRIS background spectrum.
These rates are dependent on two factors: the quality of the Gaussian fit against a baseline of the constant background ($\Delta \chi^2=\chi^2[\mathrm{linear}]-\chi^2[\mathrm{Gaussian}]$) and the integrated radiance of the Gaussian line.
Using thresholds of $I>$70 DN and $\Delta \chi^2$=0.22, we have a false positive rate of 0.02 and a true positive rate of 0.58, resulting in a likelihood ratio greater than 20.
Examples of profiles near these thresholds are shown in Figure 1.
We have a strong confidence that the line emission measured are very likely real, but we are likely eliminating many weak emission regions through our stringent detection limit.
Absolute calibration of IRIS data relies on both stellar observations and comparison of full sun spectra with the SORCE instrument \citep{rottman_05}.
The effective area of the telescope varies with time but at 1394\AA~the approximate effective area in 2014 was 1 cm$^2$.
We define radiance as the integrated spectral line photon flux at the spacecraft per subtended angle. 
Inputting our empirical detection limit, the instrument characteristics, and the exposure times, we calculate that DS3 and DS4 have a lower limit on observable line radiance of  20 ergs s$^{-1}$ cm$^{-2}$ sr$^{-1}$.
\subsection{Extraction of magnetic elements}
We use a region growing technique to segment the HMI magnetic data into magnetic elements and internetwork.
For each HMI frame, we create a set $p$ representing the index of every pixel in the image plane with an unsigned flux above 50 Mx cm$^{-2}$ (these pixels occupy less than 2\% of the quiet sun area).
We iteratively step through each pixel $p_i$ checking if any of the four abutted pixels in the image plane exceed the threshold $|B_{los}|>10$ Mx cm$^{-2}$ and each satisfactory pixel is added to the set $S_i$.
The process of connecting neighbors is continued for each element of $S_i$ until no neighboring pixel crosses the selection threshold.
Once the set $S_i$ is complete, we reduce $p$ to the set of complementary elements of $(p~\setminus~S_i)$.
This process continues until $p$ is empty.
This process must be repeated for each frame.
Once the full time series of $S$ is collected, a final filter is applied: any pixel which is not a member of $S$ for at least 6 of 11 consecutive frames or doesn't have at least 8 other members in its set is excluded.
For each image frame, we populate a 2D image ${\bf M}$
	\begin{equation}
	{{M}(i)}=\begin{cases}
	-1 & \text{where}~(i \in S)\wedge(B_{los}(i)<0)\\
	~1& \text{where} ~(i \in S)\wedge(B_{los}(i)>0)\\
	~0 & \text{where}~(i \notin S)\\
	\end{cases}
	\end{equation}
$\bf M$ defines whether a given pixel is inside a magnetic element or not.
Our weakest magnetic element in $\bf M$ has a total flux of $1\times10^{17}$ Mx.
The algorithm effectively extracts all network concentrations and many emerging/pre-coalesced concentrations \citep{parnell_02}.
Our algorithm finds about 4\% (by area) of the quiet sun is part of a magnetic element.
Synthetic raster maps are created by extracting quantities from the coaligned HMI data cubes.
\section{Si IV Radiance in the Quiet Sun}
\begin{figure*}
\centering
\includegraphics[width=.77\textwidth,trim=.5in .75in .5in .25in]{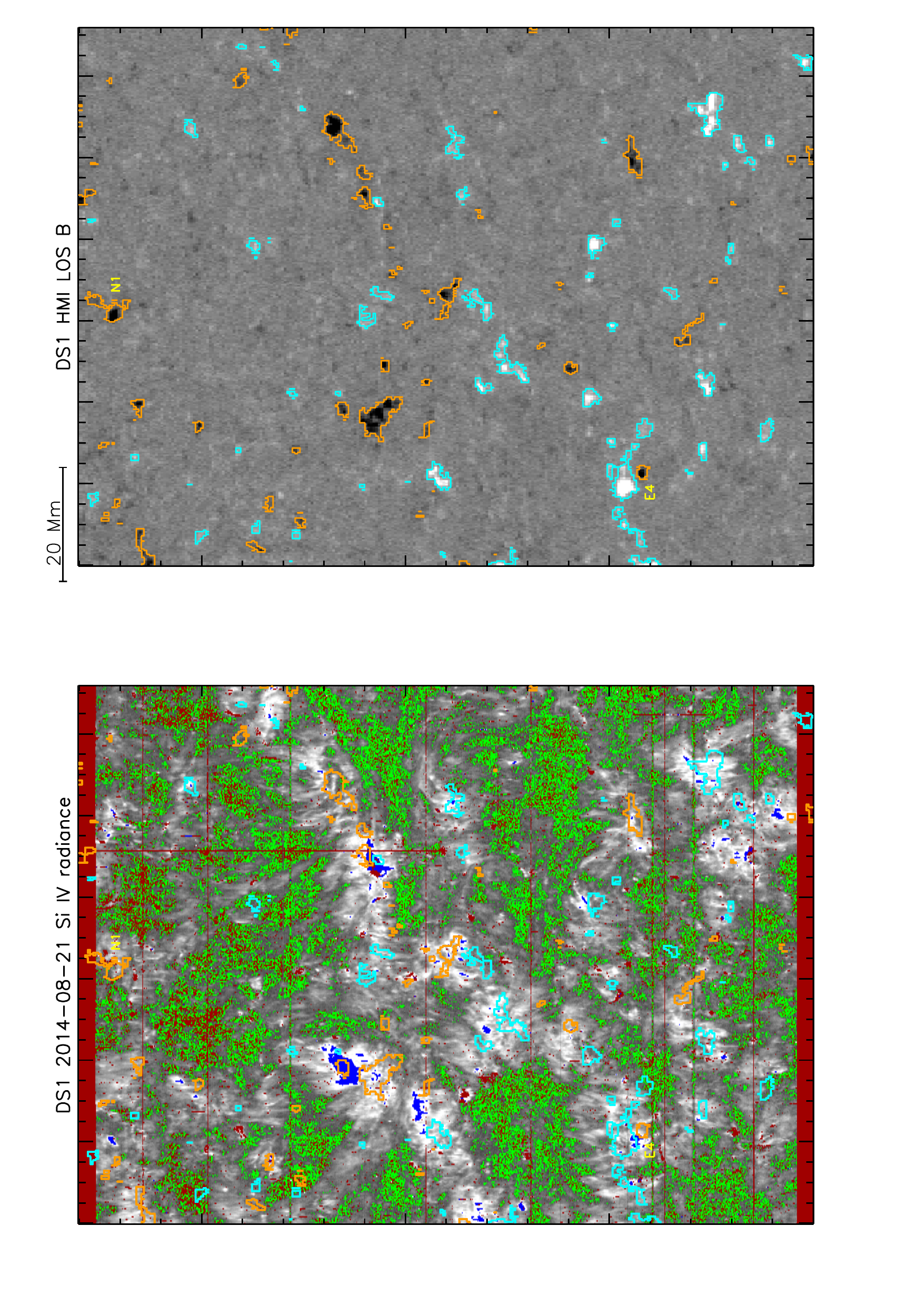}
\caption{Radiance in Si IV 1393.8\AA (bottom panel) and the aligned HMI LOS magnetograms (top panel) for DS1. Red pixels are bad data. Dark blue pixels are brighter than 1000 ergs s$^{-1}$ cm$^{-2}$ sr$^{-1}$. Green pixels have less signal than is measurable. Regions of interest are labeled with letters. Magnetic elements are outlined in orange and pale blue contours.}
\label{fig:simap}
\end{figure*}
\begin{figure*}
\centering
\includegraphics[width=.77\textwidth,trim=.5in .75in .5in .25in]{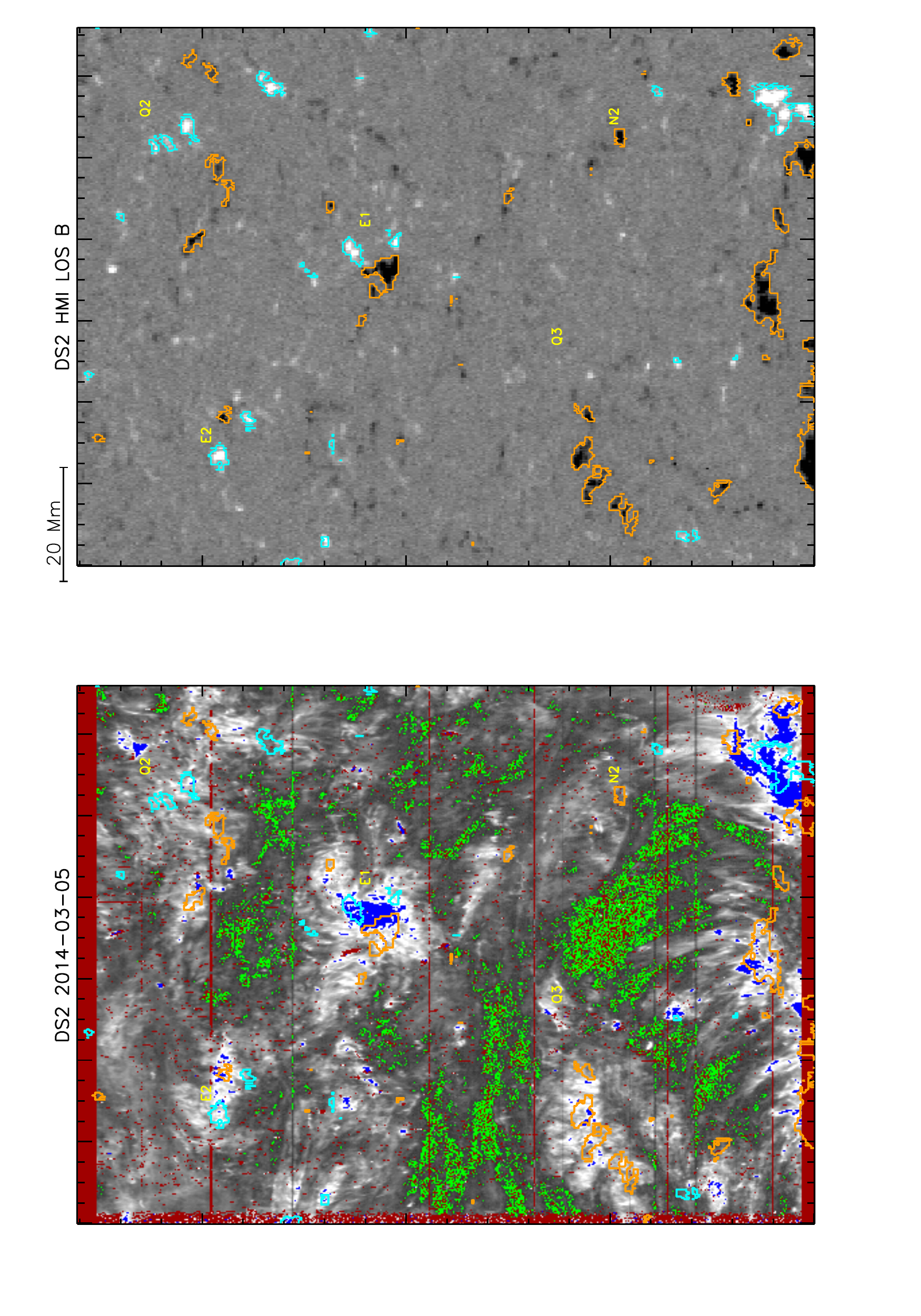}
\caption{Same as Figure \ref{fig:simap} for DS2.}
\label{fig:simap2}
\end{figure*}
\begin{figure}
\includegraphics[width=.4\textwidth]{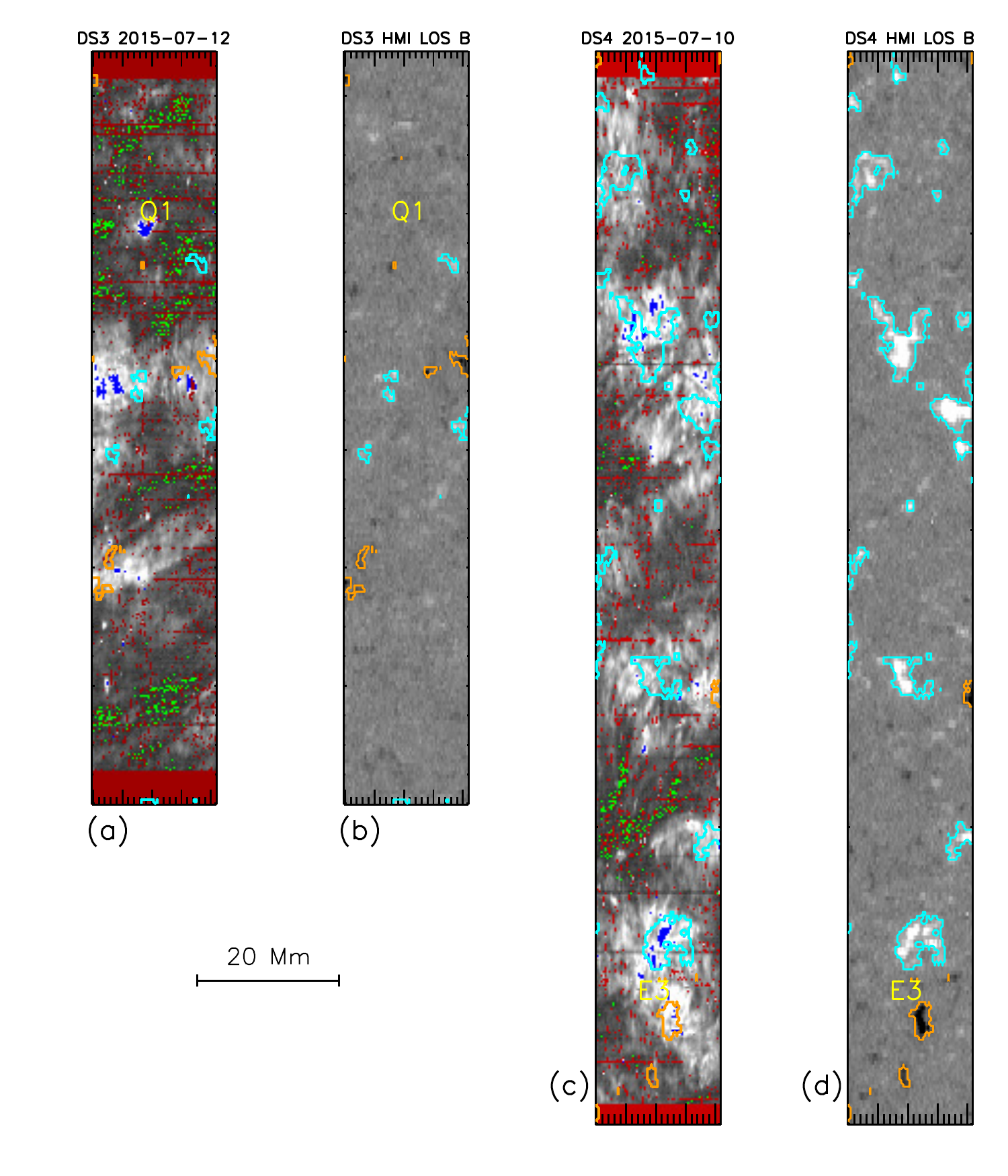}
\caption{Same as Figure \ref{fig:simap} for DS3-4.  The Si IV radiance is displayed in panels a and c and the HMI magnetogram is displayed in the panels b and d.}
\label{fig:simap3}
\end{figure}
\begin{figure}
\centering
\includegraphics[width=.5\textwidth,trim=.5cm .4cm 0 .5cm]{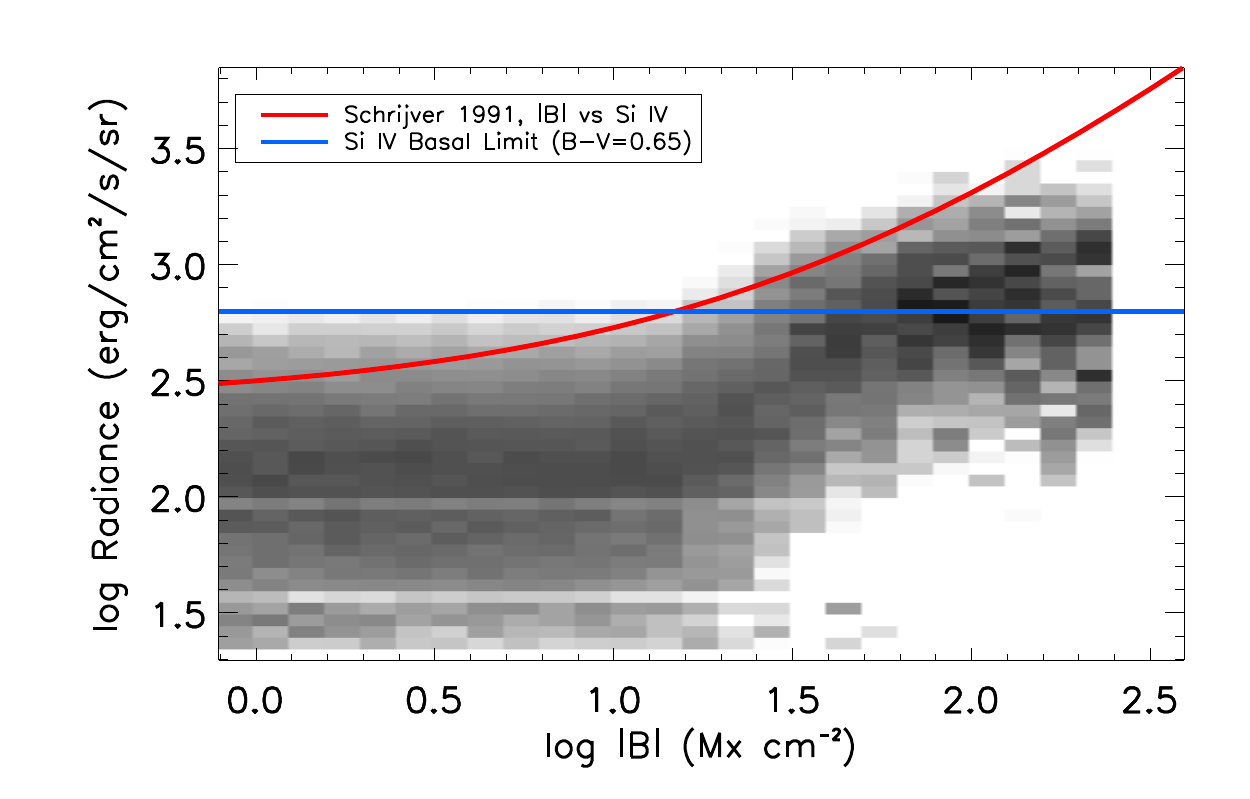}
\caption{Joint probability distribution for the line of sight magnetic field and the IRIS observed radiance in Si IV 1393\AA. The red curves shows the expected stellar relationship between $|B|$ and radiance based on a sample of stellar analogs. The blue curve shows the basal Si IV emission of Sun-like dwarf stars with low rotation rates.}
\label{fig:flxflx}
\end{figure}

The radiance in the Si IV 1393\AA~emission line as measured by IRIS is displayed in Figure \ref{fig:simap}-\ref{fig:simap3} in exponential grayscale, spanning from 20 to 1000 ergs s$^{-1}$ sr$^{-1}$ cm$^{-2}$.
As described in Section 3, the detection threshold of each dataset has been calculated and data which does not contain a strong enough signal are signified by green pixels.
To increase the contrast of the grayscale, the brightest 4\% of the observed FOV (that radiates 22\% of total emission) are shown in blue.
There are approximately 400 pixels (0.1\%) in our datasets with a radiance above 5000 ergs s$^{-1}$ sr$^{-1}$ cm$^{-2}$.\\
\indent In our deepest exposures, only 6\% of the dataset does not contain measurable emission.
There are obvious large-scale intensity differences between regions, and through visual inspection it would be reasonable to classify the observations into bright concentrations which take up a small percentage of area (enhanced network), mid-range intensities that span between the brightest regions along specific pathways (network lanes), and dim regions (internetwork).
The magnetic network is created by a dynamic process of flux emergence, advection, and accumulation \citep{parnell_02}.
Given our single-raster snapshot of the radiance structures, it is a logical leap to label the observed structures using that dynamics-based terminology.
However, the observations are consistent with the interpretation of previous authors.\\
\indent Next to each IRIS image we have plotted the simultaneous synthetic-raster magnetogram data for our four datasets.
We will go on to quantify the magnetism-radiance correlation by computing a number of magnetic statistics.
Before we delve into the statistical analysis of the data, it is important to note specific structures present in our maps.
By and large, magnetograms show a high degree of correlation with the Dowdy model.
Bright radiance structures are associated with magnetic concentrations.
In the magnetograms, we have labeled four of the most radiant regions in the data as E1-E4.
These regions occur where both positive and negative elements are present in close proximity.
This effect is easily extracted from the statistical analysis.
Two magnetic elements have been labeled N1-N2.
These elements are examples of strong magnetic concentrations which do not show any enhancement in TTP radiance.
Of the 165 well-sampled elements contained within our dataset, over 25\% do not contain pixels brighter than 500 ergs s$^{-1}$ sr$^{-1}$ cm$^{-2}$. 
These elements suggest that the creation of TTP requires more than a minimum amount of magnetic flux.
Three internetwork regions have been labeled as Q1-Q3.
These are examples of bright emission that are not located in close proximity to a magnetic element.
These suggest that weak fields can create TTP.\\
\indent The connection between magnetic fields and enhancement in radiative losses has been well established in both chromospheric \citep{skumanich_75} and coronal emission \citep{vilhu_84,reeves_76}.
Figure \ref{fig:flxflx} plots a joint occurrence distribution of magnetic flux and radiance in a so-called flux-flux plot similar to \citet{schrijver_89}.
The distribution is a broken power law with an elbow at $|B|$=20 Mx cm$^{-2}$.
For $|B|<$20 Mx cm$^{-2}$, the radiance is not well correlated with the magnetic field.
This may be due to the limited sensitivity of the HMI dataset.
For $|B|>$20 Mx cm$^{-2}$, there is a positive trend (slope$\approx 1.5$) between these variables.
The variation in radiance is significantly lower at high $|B|$ than low $|B|$.\\
\indent The colored lines in Figure \ref{fig:flxflx} are based on previous solar/stellar measurements.
Extending the work of \citet{schrijver_89}, subsequent authors compared chromospheric fluxes with coronal and x-ray fluxes across large samples of stellar types \citep{rutten_91,zwaan_91, schrijver_91}.
These seminal papers for solar-stellar comparison juggled the available measurements to establish a common metric.
The most pervasive measurement, flux in the Ca II chromospheric lines, was found to be well correlated with stellar color and TTP emission as well as solar resolved magnetic flux density (after accounting for a basal flux).
The red curve plotted in Figure \ref{fig:flxflx} was derived via the comparison of the stellar Ca II flux-Si IV flux trend and the solar $|B|$-Ca II flux trend.
At low field strength, the Ca II data (and our Si IV data) approach a minimum level of emission, termed the basal flux.
The solar Ca II data suggested there was atmospheric heating not associated with resolved magnetic concentrations.
By tapping into stellar data, specifically using samples spanning stellar types and rotation rates, it was found that Main Sequence dwarf and giant stars exhibited a minimum amount of TTP emission.
Stellar rotation is thought to power the dynamo action that generates large-scale magnetic activity in stars \citep{noyes_84}, so chromospheric emission from stars outside the dynamo regime suggests that local convectively-driven dynamo action or alternatively purely acoustic heating must play a role.
The horizontal blue line shows the minimum detected limit of Si IV emission for $B-V=0.65$ stars based on IUE data \citep{rutten_91}. 
The Sun is believed to be a relatively inactive star for its age and spectral type so it is not surprising that much of our dataset sits below the basal limit.\\
\indent One of the basic takeaways from Figure \ref{fig:flxflx} is that there is significant spread in radiance per magnetic bin.
This is especially true for weak field regions.
In order to extract more information for our physical interpretation, we will further investigate the magnetism-radiance relationship by incorporating spatial information as well.
The photospheric magnetic field is not uniformly distributed in space and strength. 
Rather equiparition flux tubes coalesce into larger more stable structures, clumps of magnetic field that are unipolar over a few square megameters.
\begin{figure*}
\centering
\includegraphics[width=.85\textwidth]{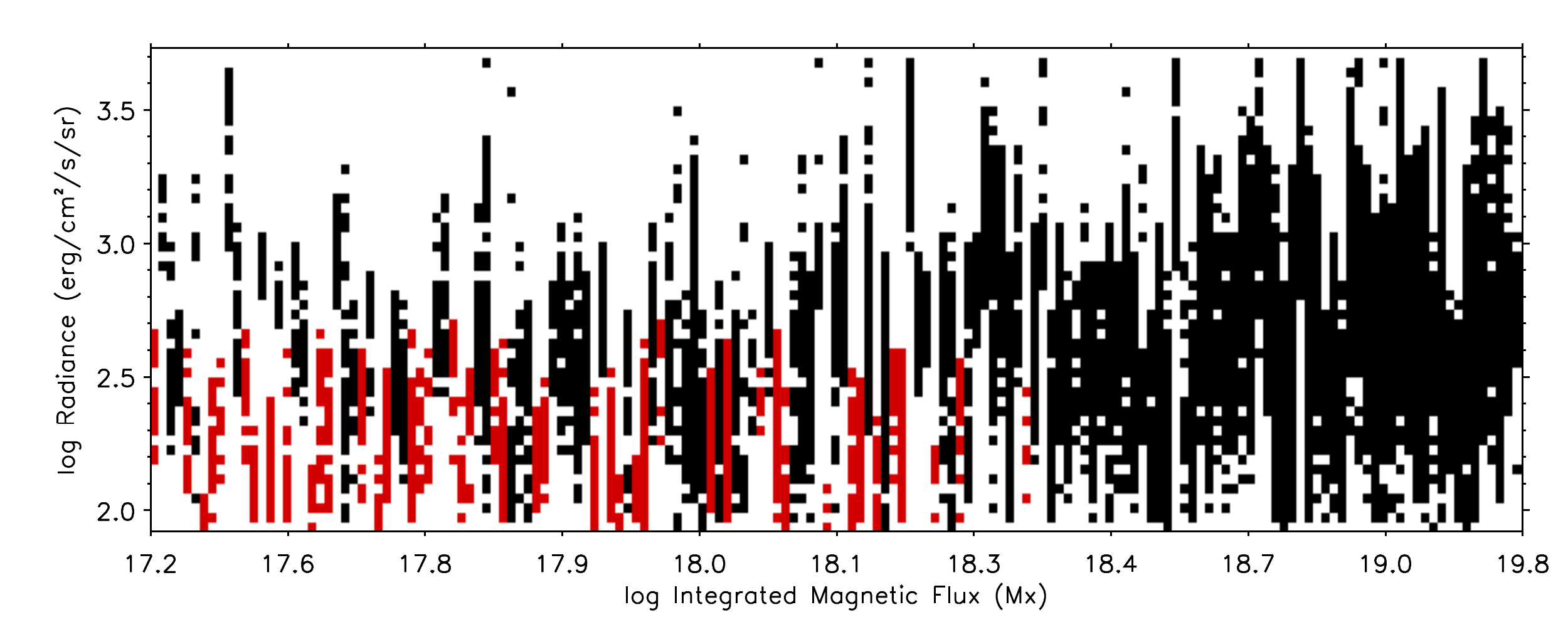}
\caption{Distribution of radiance per magnetic element. Black or red blocks denote that at least one pixel overlying that element emits at that radiance. Red columns denote elements that are not brighter than the network lane background.}
\label{fig:flxrad}
\end{figure*}
These clumps represent the basic structural unit for the photospheric network.
Figure \ref{fig:flxrad} illustrates the distribution of intensities we observe over each individual magnetic flux concentration.
Each column present a binary histogram indicating if any pixels overlying that magnetic element radiate at that intensity.
We have data for 165 elements for which we have at least 9 pixels of IRIS spectrograph overlap.
Given the trend in Figure \ref{fig:flxflx}, we would expect that large flux concentrations would be concentrated at higher radiances than small flux concentrations.
While larger concentrations often do extend to higher intensities, we find that most elements also contain dim regions well below the 500 erg cm$^{-2}$ s$^{-1}$ sr$^{-1}$ threshold (95\% of the internetwork radiates at lower levels).
The presence of this dim emission has not be previously discussed in analysis of network TPP structure.
For smaller flux concentrations we find several instances where elements contain highly discontinuous radiance distributions, with a few high intensity outliers.
This provides ancillary evidence that thermal energy is deposited in the network inhomogeniusly.
Spicules must be at least partially responsible for the patchy bright emission.
We know that they dynamically produce TTP emission (with a 100 second lifespan) and they exhibit sub-structure down to at least 0.05" scales \citep{pereira_14}.
We do not know their filling factor.\\
\indent One last aspect of Figure \ref{fig:flxrad} should be mentioned.
Approximately 25\% of the observed elements do not have any pixels that emit above the nominal internetwork threshold (columns in red, see paragraph above).
It is obvious that the presence of strong flux (each element has at least one pixel where $|B| >50$ Mx  cm$^{-2}$) does not by itself create enhanced TTP emission.\\
\indent Figure \ref{fig:flxrad} provides us some spatial context for the magnetism-radiance relationship: how does the size of magnetic concentrations affect emission?
This diagnostic is only valid for regions that directly overlie magnetic elements.
We wish to extend it by considering how the proximity to magnetic elements affects the surrounding internetwork.
As an example, consider the geometry of network funnels in the Dowdy model.
Horizontal segments of loops will project onto the internetwork when viewed from above.
Consider two highly simplified processes that produce TPP: conductive back-heating from the corona and dynamic in situ heating.
Hypothetically, these processes could occur along identical funnel shaped flux tubes.
Under back heating conditions (i.e.  hydrostatic coronal loops), the extent of TTP emission will be determined by the altitude at which the plasma temperature falls within the TTP range and the angle of funnel expansion at that altitude.
With dynamic in situ heating, there is no static TTP.
TTP is formed when energy is deposited in the chromosphere: some energy is radiated, some energy is thermalized, and some energy accelerates plasma.
The extent of TPP emission would be determined by the density and velocity of the TTP (or plasma that subsequently cools into the TTP range) as well as the geometry of the funnel.\\
\indent Magnetic proximity is defined as the radial distance in the image plane between each pixel and the nearest non-zero pixel in {\bf M} for each magnetic polarity
\begin{equation}
R^+_{i}=\text{min}( |R_{i}-{\bf R}|) \in {\bf M}=1,
\end{equation}
and
\begin{equation}
R^-_{i}=\text{min}( |R_{i}-{\bf R}|) \in {\bf M}=-1,
\end{equation}
where $R_{i}$ is the value of element $i$ in matrix {\bf R} and ${\bf R}$ is the position of each pixel in Cartesian coordinates.
Once {\bf R$^+$} and {\bf R$^-$} are determined, we further sort the distance:
\begin{equation}
{\bf r^0}=\text{min}({\bf R^+},{\bf R^-})
\end{equation}
and
\begin{equation}
{\bf r^1}=\text{max}({\bf R^+},{\bf R^-})
\end{equation}
so that the polarity information is discarded.
Our proximity statistics, $r^0$ and $r^1$,  represent the Cartesian distance in the image plane from the IRIS observation to the position of the nearest elements of both polarities, where $r^0<r^1$.
Low-$r^0$ values represent pixels nearby to magnetic elements, while high-$r^0$ values represent regions like cell-center internetwork.
Low-$r^1$ values represent pixels that are nearby to both polarities, while high-$r^1$ values occur in large-scale unipolar regions.
The mean $r^0$ value of our datasets is 4 Mm and the mean of $r^1$ is 8.5 Mm.\\
\indent Figure \ref{fig:magdep} displays the effect magnetic proximity has on TTP radiance.
For  Figure \ref{fig:magdep}a we only consider the relationship between $r^0$  and radiance, and we ignore $r^1$.
The bar for each $r^0$ value is derived by calculating the width of the log normal distribution as described by \cite{griffiths_99}, while the diamond marks the distribution centroid.
Radiance is highest near elements (as we would expect). 
An approximate powerlaw fit has a slope of -0.6 although it is not clear that a powerlaw relationship is a reasonable assumption.
The convex tail at high $r^0$ is statistically significant; our dataset has approximately $4\times10^4$ pixels for $r^0>7$ Mm.
We suggest the high-$r^0$ bump in the radiance-$r^0$ relationship is an indication of emission from internetwork-internetwork connected loops. 
To determine the robustness of this analysis, let us consider two potential biases.
First, given an inhomogeneous (clumpy) but isotropic (independent of location) distribution of magnetic concentrations on the photosphere, does our magnetic element detection threshold produce a bias effect for large $r^0$?
We have examined the distribution of unsigned and sub-detection level flux density in our dataset.
For $r^0$=5 Mm to $r^0$=12 Mm, we find that the mean unsigned flux density varies at the 5\% level around 6 Mx cm$^{-2}$ without a trend.
If our magnetic detection algorithm were skewing the measurements such that the most distant internetwork was falsely identified (i.e. it lies nearby to undetected elements), we would measure an increase in the mean unsigned flux density.
Second, does our calculation of $r^0$ predispose the radiance-$r^0$ relationship to contain a minimum point?
This insidious bias is best considered geometrically.
\begin{figure*}
\centering
\includegraphics[width=.85\textwidth]{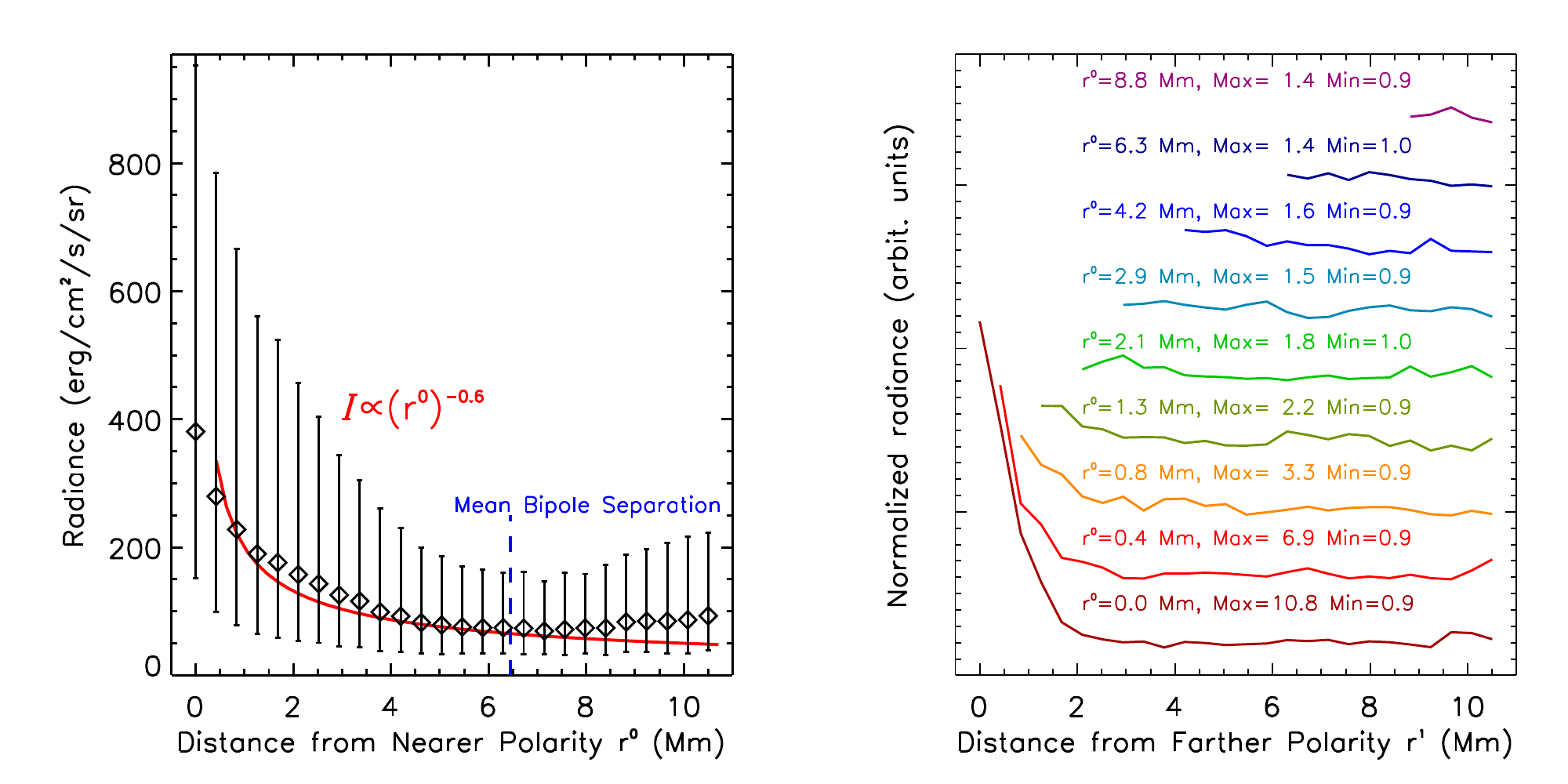}
\caption{The effects of magnetic proximity on Si IV radiance.(a) We marginalize (i.e ignore) ${r^1}$. The vertical bars are the 1$\sigma$ limits of the log-normal distribution. (b) We control for ${r^0}$ and only test the trend between $r^1$ and radiance. Each curves represents a isovalue of $r^0$. A y-axis offset has been added to each curve.}
\label{fig:magdep}
\end{figure*}
Our $r^0$ statistic asks which element is closest to the observation and how far away is it.
Given the isotropic distribution of elements in the photosphere, it is statistically more likely that as you enlarge the circle where you check for elements, you are likely approaching more than one element (this is similar to a packing problem in geometry).
Can this effect produce a bump in the $r^0$-radiance relationship?
We tested this effect by considering a 200 Mm x200 Mm patch of quiet sun (see potential field in Section 5).
We extracted the position of 250 magnetic elements and we synthesis TTP emission by considering that each element contributed to the emission at any given point according to its distance from that point.
We implemented both an exponential and a powerlaw radial emission model.
Using the synthetic data, we compared the statistical relationship for radiance (with summed contributions from all elements) and the $r^0$ statistic which relates each pixel with only a single (the nearest) element.
We found that neither the powerlaw nor the exponential model produced a bump at high $r^0$.\\
\indent In addition to the $r^0$ statistic, we have also parsed the observations using the $r^1$ statistic. 
The $r^1$ statistic relays information on how the presence of mixed polarity fields affect radiance.
To isolate the effect that $r^1$ has on radiance, we isolate isovalues of $r^0$ (0.4 Mm bins) and calculate the median radiance:
\begin{equation}
\langle I(r_a, r_b)\rangle=\frac{1}{N}\sum^N I({\bf r^0}=r_a,{\bf r^1}=r_b)
\end{equation}
where $N$ is the number of points in the set $({\bf r^0}=r_a,{\bf r^1}=r_b)$.
Figure \ref{fig:magdep}b plots the normalized radiance, $\langle I(r_a, r_b)\rangle'$,  which has been scaled by the median of the radiance marginalized for $r^0$ (shown as diamonds in Figure \ref{fig:magdep}a):
\begin{displaymath}
\langle I(r_a, r_b)\rangle' =\frac{\langle I(r_a, r_b)\rangle}{\langle I (r_a)\rangle}.
\end{displaymath}
The derived curves $\langle I ({\bf r^0}=r_a,{\bf r^1})\rangle'$ are plotted in  Figure \ref{fig:magdep}b with an indication of the minimum and maximum values of the curve.
\begin{figure}
\centering
\includegraphics[width=.5\textwidth]{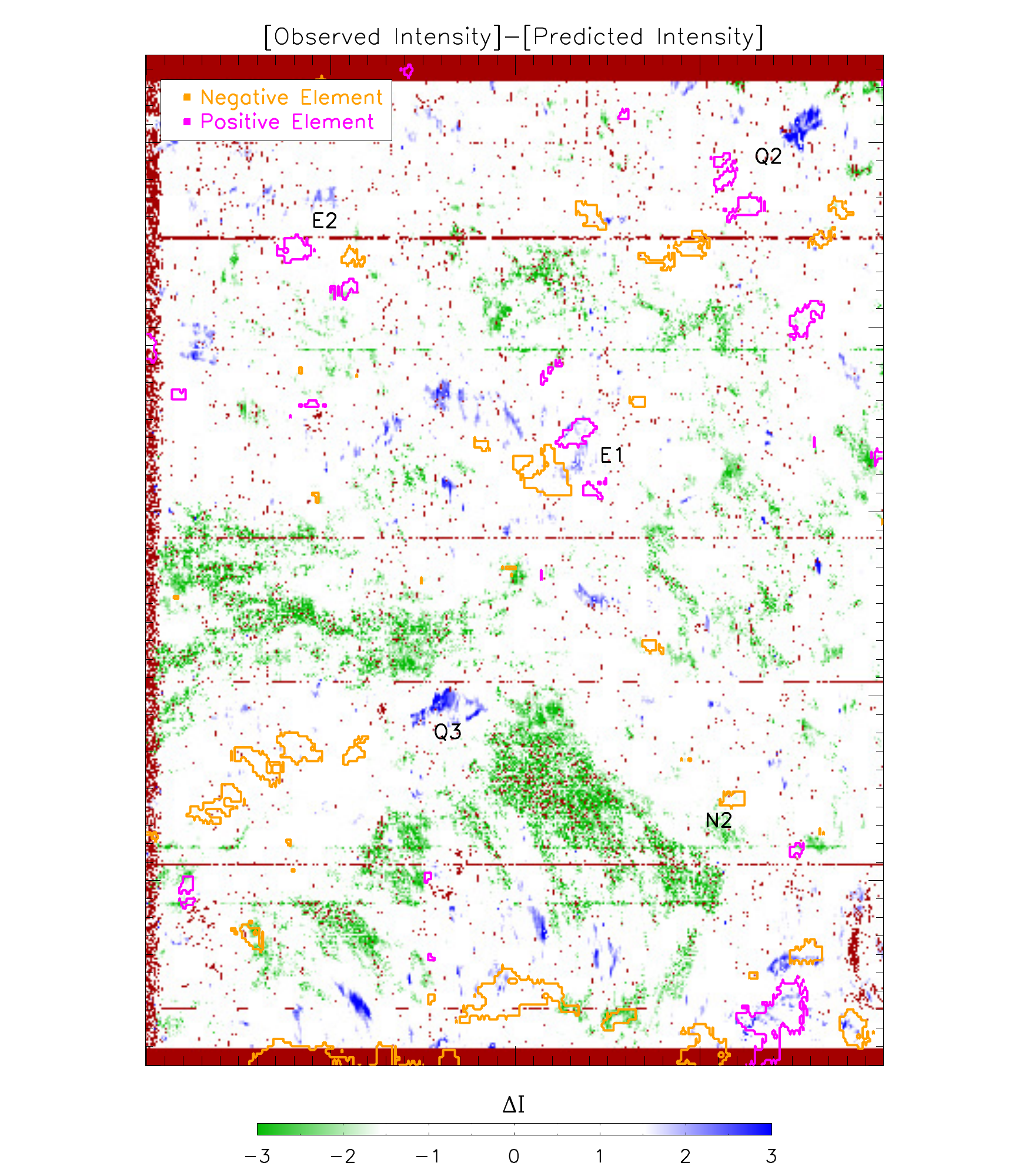}
\caption{Magnetic residual as calculated by Equation \ref{equ:res} for DS2. Our magnetic proximity statistic is unable to differentiate between network lanes and cell interior brightness regions. Orange and pink contours show magnetic elements.}
\label{fig:std}
\end{figure}
Consider a model where short-span loops have no variation in emission measure over their length (like the UFS reported by \citealt{hansteen_14}), but long-span loops have a radial falloff with $r^0$.
The resulting radiance-$r^1$ relationship will have two components: high radiance at low $r^0$, low $r^1$ and low radiance at high $r^1$.
That model is one simple example how connectivity and $r^1$ can be related to emission structures.\\
\indent There are two trends in Figure \ref{fig:magdep}b of importance.
First, for $r^0<2$ Mm and $r^1< 3$ Mm we find that there is a negative trend that links the proximity of opposite polarity field to enhancement in radiance.
As an example, a pixel that overlies a positive magnetic element and also sits within 1.2 Mm of a negative polarity element is a factor of 2.7 brighter than if that same pixel were 8 Mm from a negative polarity element (on average).
While this effect can be drastic, it disappears rapidly.
For $r^0>2$ Mm and for  $r^0<2$ Mm and $r^1>4$ Mm, the trends are more or less flat.
This implies that the link between mixed polarity concentrations and enhanced emission are limited to short scales.
We suggest that this sets a limit on the length of network-connected UFS loops on average.\\
\indent  Now that we have an expectation value, $\langle I(r_a, r_b)\rangle$, that prescribes how intensely each pixel should emit based on its magnetic proximity, we will create a statistic, magnetic residual, that allows us to examine how outliers (pixels emitting far different amounts of radiation as compared to the predicted value given its magnetic environment) are spatially distributed in our maps.
The magnetic residual is calculated as:
\begin{equation}
\Delta I(r_a,r_b)= \frac{I(r_a,r_b)-\left<I({\bf r^0}=r_a,{\bf r^1}=r_b)\right>}{\sigma({\bf r^0}=r_a)}.
\label{equ:res}
\end{equation}
where $\sigma({r^0})$ is the log-normal gaussian width as depicted in Figure \ref{fig:magdep}a.
The magnetic residual map for D2 is plotted in Figure \ref{fig:std}.
The strongest positive difference (brighter than expected, shown in blue) tend to occur in small (10-20 Mm$^2$) clumps.
They occur both nearby and far removed from magnetic concentrations.
Q2 and Q3 mark two particularly strong emission areas that do not obviously connect to elements.
The general appearance of the negative residual regions (dimmer, shown in green) are more contiguous and larger scale ($>200$ Mm$^2$) .
Using the standard network/supergranule terminology, these are cell interior regions.
\begin{figure*}
\centering
\includegraphics[width=.8\textwidth]{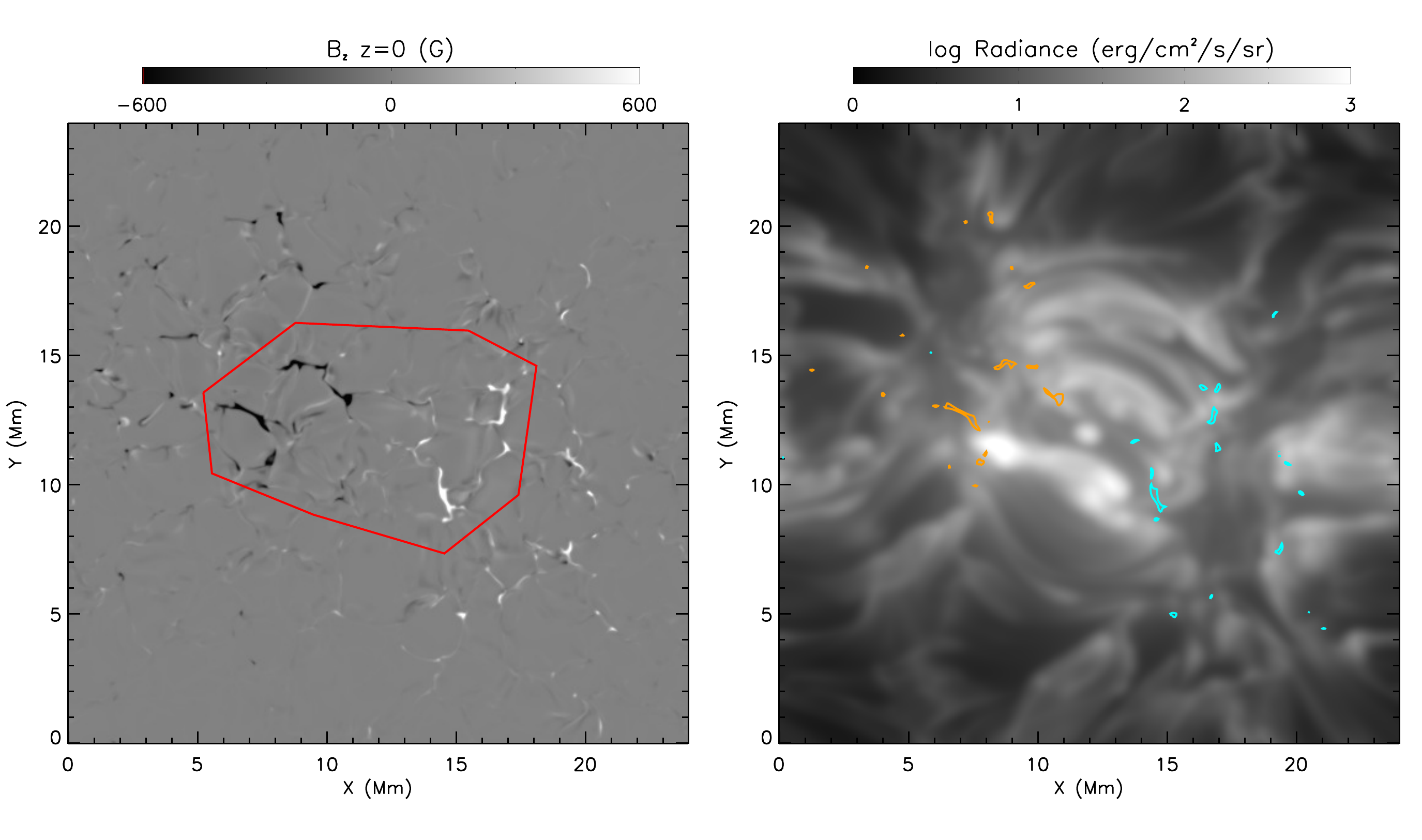}
\caption{Snapshot from the Bifrost simulation at t=5030s. The vertical magnetic field at z=0 (a). The red polygon borders the bipole region. Synthesized Si IV 1393\AA~emission convolved with the IRIS point spread function. Magnetic concentrations are outlined in blue and orange. }
\label{fig:bif2d}
\end{figure*}
\begin{figure}
\centering
\includegraphics[width=.5\textwidth]{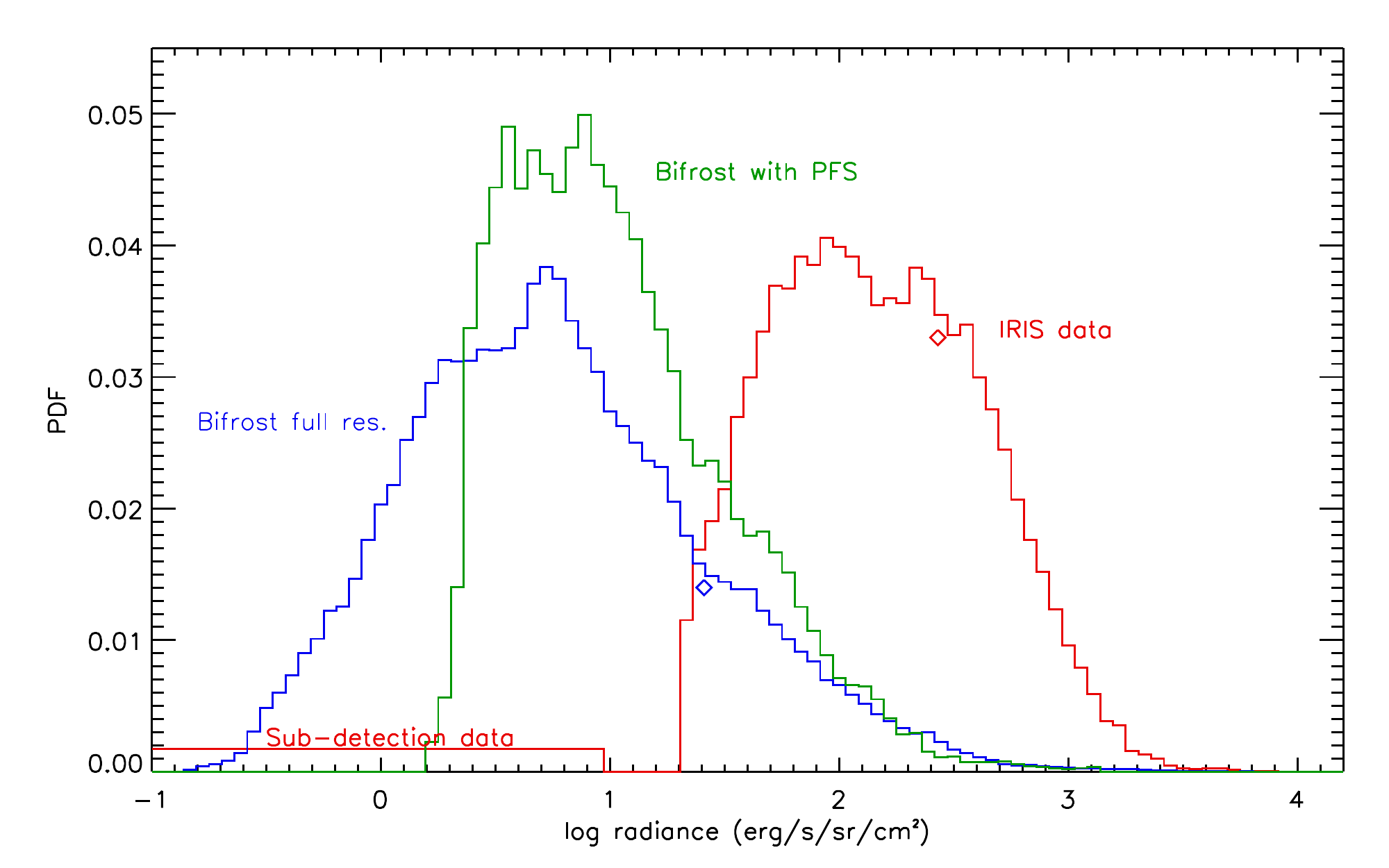}
\caption{Comparison of log-normal radiance probability distribution function (PDF i.e. occurrence rate) in the Bifrost model with IRIS DS3 and DS4. Diamonds mark the mean radiance of the Bifrost and IRIS dataset. The green curve has convolved the Bifrost output with the IRIS point spread function.}
\label{fig:bifdist}
\end{figure}
We have found that we are unable to recover the respective location of network lanes and cell interiors using magnetic proximity.
Our initial assessment of TTP spectroheliograms is accurate: the magnetic network is bright.
However, the diffuse and moderate emission we associate with network lanes often occurs at $r^0>$ 8 Mm which is identical to many patches of dim cell interior emission.
This implies that our magnetic proximity statistic ignores some unknown basis quantity that differentiates the cell interior from the network lane.
\section{Magnetic Structure in the Quiet Sun}
In the previous section, we have used photospheric magnetic information to parse the IRIS SI IV data.
Overall, we are interesting in linking the TTP emission structures with magnetic connectivity.
Toward that aim, we will use magnetic models to estimate how different loop types emit in the TTP regime and how different loop types are partitioned in the sub-coronal atmosphere.
\subsection{Bifrost MHD model}
The magnetic field above the photosphere evolves with the turbulent motions of convection cells.
This process energizes the field and it is believed that at least a fraction of that energy is thermalized to heat the chromosphere and corona.
While loop models \citep{reale_10} can be constructed to study the static or dynamic conditions of plasma isolated to a 1D structure, these models are at best an approximation of the linked evolution of magnetic and thermal conditions.
3D MHD codes act as a numerical experiment in which these quantities can be studied self-consistently.
In order to form a chromosphere in such a model, non-grey radiative transfer must be taken into account.
The Bifrost code \citep{gudiksen_11} is the state-of-the-art in these calculations.
The Bifrost team has made some simulation data freely available as described in \cite{carlsson_16}.
The simulation models the emergence of an enhanced network region.
The magnetic model initializes with a bipolar magnetic distribution specified at a depth of 3 Mm below the photosphere at t=1750 seconds (in solar time, where t=0s is the initialization of model without field).
We analyze a single timestep of the simulation (t=5030s).
\begin{figure*}
\centering
\includegraphics[width=.9\textwidth]{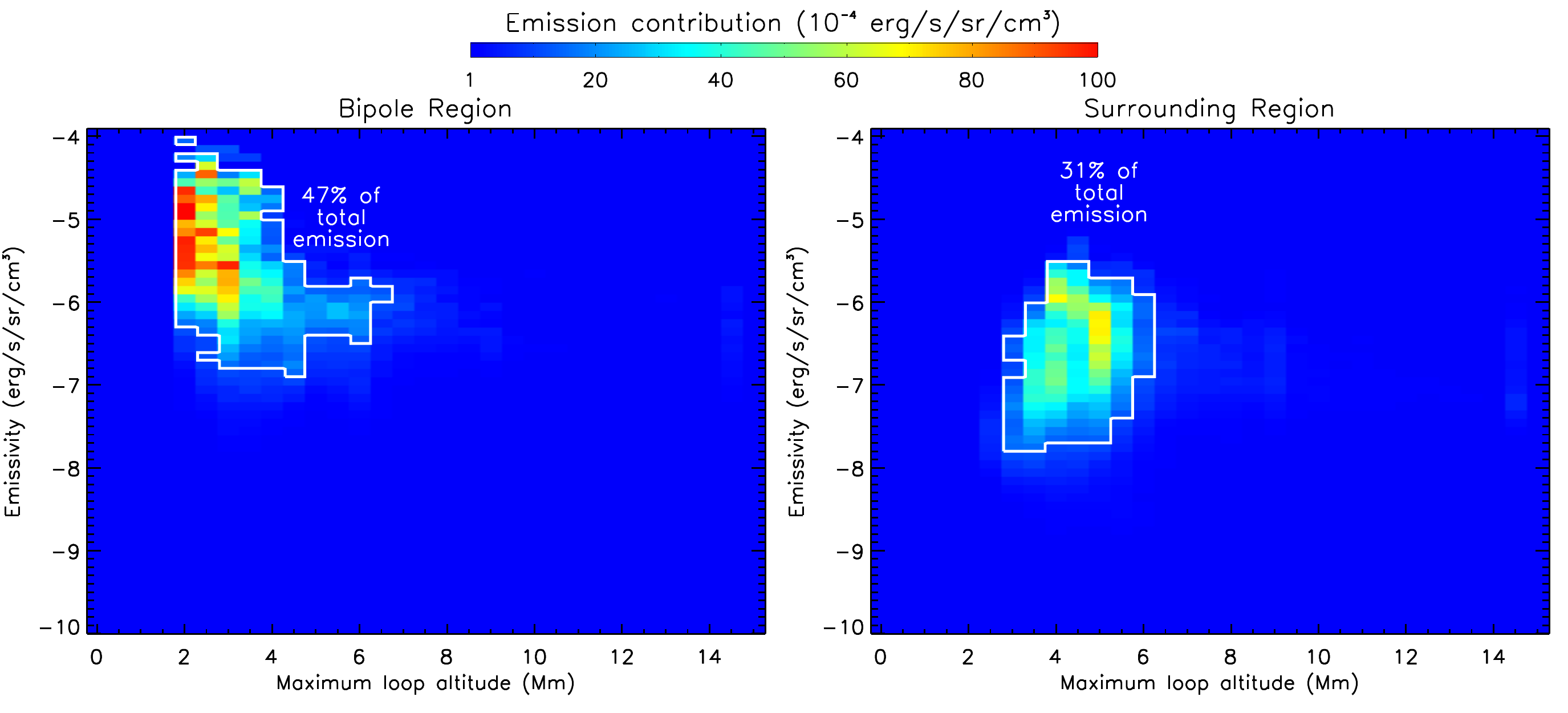}
\caption{Contribution to total volume emissivity (1393\AA) of the Bifrost simulation, partitioned by max$[h]$ and emissivity per grid element.}
\label{fig:bifh}
\end{figure*}
At this timestep, convective motions have had time to bring magnetic flux up to the photosphere, and magnetic loops span the chromosphere and corona, connecting the bipolar flux concentrations.
We will use Bifrost to examine where emission occurs in the simulation and which type of magnetic loops are associated with TTP plasma.\\
\indent Figure \ref{fig:bif2d} shows the photospheric magnetic field and synthesized emission in Si IV 1393\AA.
The emerged magnetic structures exhibit a continuum of fluxes and areal scales, with the largest elements covering 8 Mm$^2$ .
The emerged structure is largely bipolar with approximately 10 Mm separating the polarities.
Looking away from the bipole region, we see that the simulation lacks a significant amount of the weak internetwork fields that are observed on the Sun \citep{orozco_07}.
The Si IV emission in Figure \ref{fig:bif2d} has been convolved with a point spread function (PSF) that mimics that of IRIS.
Two features jump out of the synthetic image.
Firstly, we find that dim and bright loops often neighbor each other with a characteristic width (cross section) of 0.5-2 Mm.
Secondly, the brightest features in the image are nearby but not co-spatial with magnetic elements.
The first feature is consistent with the IRIS data.
As was illustrated in Figure \ref{fig:flxrad}, dim and bright structures are mixed over small scales.
The second feature is not completely consistent with the IRIS data. 
While we find that many magnetic elements contain dim regions, on average we find that magnetic elements are brighter than the pixels that surround them.
This is backed by the flux-flux relationship.
We suggest that the effect seen in the simulation is at least partially due to strong emission that occurs  along relatively horizontal field lines where projection effects generate the offset.
\\
\indent Outside the bipole region, mid-intensity loops are visible that seemingly end at the simulation box borders.
The simulation is periodic in the x- and y-direction and these loops mostly connect the weaker magnetic elements outside the bipole region.
The magnetic connectivity of the simulation is simpler than quiet sun, but the distributed flux and periodic boundaries help to create a more complex magnetic environment than an initial inspection might suggest.\\
\indent Figure \ref{fig:bifdist} shows the direct comparison between the distribution of synthesized 1393\AA~radiance and our two deepest IRIS datasets (DS3 and DS4).
\cite{olluri_15} were the first to analyze optically thin emission from TTP with Bifrost.
We use identical algorithms and line synthesis parameters  as those authors to forward model emission.
The simulations have similar magnetic setups although the publicly available cube calculates non-LTE hydrogen ionization in the energy equation.
We have found that in \cite{olluri_15} the SUMER quiet sun radiance values were misrepresented (V. Hansteen, private communication).
As displayed in Figure \ref{fig:bifdist}, Bifrost produces too little emission in 1393\AA~by approximately a factor of 10 at the mean.
At full resolution, the Bifrost simulation produces a wider distribution of radiances and exhibits an enhanced high-radiance tail.
By incorporating an IRIS-like point spread function into the data, we find that the median intensity is shifted higher by approximately a factor of two, but the distribution still differs significantly from the IRIS data.
Based on the synthesis of optically thick lines, it has been documented that the densities in the upper chromosphere of Bifrost are too low \citep{rathore_15}.
Our analysis indicates that this discrepancy extends to TTP
Previous authors have also discussed how Bifrost does not produce spicules with regularity \citep{jms_13}.
The connection between spicules and TTP density will be discussed in Section 6.
While Bifrost is not a perfect match to the solar atmosphere, it is a numerical experiment meant to mimic the Sun by including the physics we believe to important.
In particular, it provides the most detailed testbed to understand the joint evolution of magnetic fields and plasma thermodynamics through the $\beta=1$ layer.\\
\indent Magnetic geometry is a fundamental quantity in the energy balance in the transition region and corona.
For each point in the Bifrost atmosphere, we have drawn field lines according to the relation
\begin{displaymath}
d\vec{s}=\frac{\vec{B}}{|B|}
\end{displaymath}
and each model grid cell is assigned the value of the maximum altitude, max[$h$], achieved by the embedded field line.
In Figure \ref{fig:bifh}, we compare the 3D max[$h$] field with the 3D distribution of emissivity (radiance per unit length along the line of sight).
The color scale denotes the total contribution to emission (of the volume) from that histogram bin.
The simulation volume has been segmented into two regions (as shown in Figure \ref{fig:bif2d}): the bipole region (inside the red polygon, 15\% of total volume) and the surrounding area.
We find that in both regions the emission is dominated by short loops although the relative emissivity is important.
There is a broad range of emissivities that contribute at a high percentage level to total emission: from 10$^{-4}$ (in the bipole) to 10$^{-8}$ erg s$^{-1}$sr$^{-1}$cm$^{-3}$.
Accounting for the log-scaled axis, this means that for every single grid cell that emits at 10$^{-4}$ erg s$^{-1}$sr$^{-1}$cm$^{-3}$ there must be 10$^4$ grid cells that emit at 10$^{-8}$ erg s$^{-1}$sr$^{-1}$cm$^{-3}$ to generate an equivalent emission contribution.
If this were attributable to a temperature effect (emission in the wings of $G(T)$), it correlates with plasma at $3\times10^4$ K or $6\times10^5$ K.
If this emission were attributable to a density effect, a variation in $n$ of 10$^2$ could generate the diffuse emission.
Approximately 95\% of the emission in the simulations comes from two scale heights in altitude, 1.5 Mm$<z<$4.5 Mm (weighted 3:1 towards the lower half), which we refer to as the TPZ (Transition Plasma Z-range).
This implies that there are significant differences loop-to-loop that generate such disparate plasma conditions within such a narrow swath of atmosphere.\\
\indent High altitude loops, our stand in for coronal structures, contribute a surprisingly small amount to total emission.
Overall, loops where max$[h]>$7 Mm contribute less than 9\% of the total emission of the simulation.
If we look only at the bipole region, we find that tall loops comprise only 6\% of the regional emission despite a volumetric filling factor of 27\%.
There is a caveat that we must acknowledge in our interpretation of the model.
The simulation boundary conditions likely affect loops that reach the upper boundary of the box.
We believe that the loop properties of the model are well constrained only for loops less than 14 Mm in height and less than 50 Mm in arclength.
\begin{figure*}
\centering
\includegraphics[width=\textwidth]{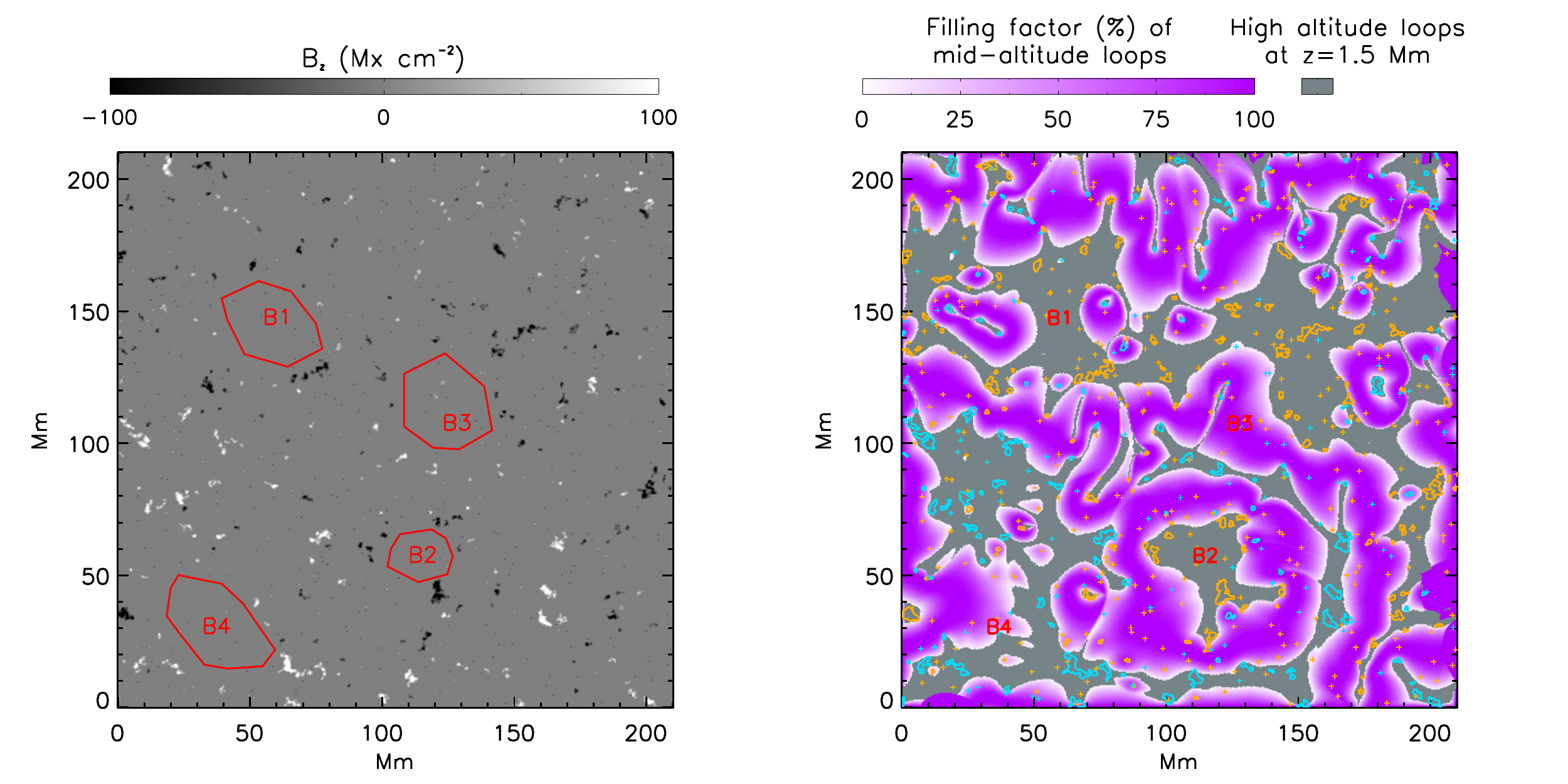}
\caption{Vertical magnetic field from processed HMI data for a flux-balanced quiet sun region (a). The red polygons show example cell interior regions.The filling factor of mid- and high-altitude loops in the TPZ region of the field model (b). Magnetic elements are shown in blue and orange contours. The additional weak field point sources are indicated by crosses. }
\label{fig:pfss}
\end{figure*}
Our interpretation of the Bifrost emission structure can be summarized as follows.
Si IV emission occurs over a narrow range of altitude (the TPZ).
Short loop emission dominates over tall loop emission.
The brightest short loops are isolated to regions connecting large flux concentrations.
The plasma conditions can vary drastically at the pertinent altitudes but diffuse and wide-spread emission can make a significant contribution to the integrated field radiance.
When considering these results, we must also consider the limitations of the model.
It does not produce spicules, and it produces significantly less TTP emission than the Sun.
\subsection{Potential Field Model}
\indent In its current implementation, Bifrost cannot calculate the temporal evolution of the large-scale network.
The length scales (at least 50 Mm on a side) and timescales (days or longer) are too computationally expensive.
If we are interested in understanding how different loop geometries are distributed throughout a diverse magnetic landscape more akin to real quiet sun, force-free models remain the only option.
Potential fields are an inherent simplification of the chromospheric and coronal magnetic field.
No currents are present in a potential field, although we know that currents are likely common in the solar atmosphere.
The equipartition layer between thermal and magnetic pressure resides in the chromosphere, and dynamic motions of the plasma can do work on the magnetic field.
We will not be using a potential field to try to match specific loop structures, rather we will use the model to analyze loop statistics based on an observed photospheric magnetic field.
This model represents a magnetic volume that is the most simplistic and contains the least magnetic energy possible (for the given boundary condition).\\
\indent We calculate a potential field model of a large swath of quiet sun having made the following considerations.
First, we use an algorithm to find a region of the solar disk that is flux balanced, working from a 45s-cadence HMI magnetogram dataset on 21-Aug-2014.
We identified a well suited 210 Mm x 210 Mm area just off disk center.
Similar to our IRIS analysis, we isolate magnetic elements with an integrated flux greater 3$\times10^{17}$ Mx.
All regions outside the elements are zeroed to eliminate spurious magnetic features.
The resulting magnetic distribution is unbalanced with a net positive imbalance of 1.5\% and an unsigned flux of 7.3$\times10^{20}$ Mx.
The fluxes are balanced through the insertion of 274 randomly distributed point sources of 50 Mx cm$^{-2}$ each.
Additionally to mimic the presence of unresolved weak fields, we insert 200 point sources of balanced polarity.
These modification steps are intended to remove any spurious instrumental effects from our dataset while maintaining our ability to analyze the connectivity of weak (non-network) fields within a controlled environment.
This magnetic field map is used as the vertical field boundary condition for a potential field extrapolation computed using the FFF routine which is including in the NLFF library of Solarsoft.\\
\indent Figure \ref{fig:pfss}a shows the magnetic field map we use as the boundary condition for our extrapolation.
A potential field is calculated in the $\beta=0$ limit so we cannot synthesize TTP emission as we did with the MHD model.
We can however apply the understanding gained from the Bifrost model to our new magnetic landscape.
In particular we can calculate how mid-altitude loops are distributed within the volume and how that distribution relates to magnetic concentrations.
Figure \ref{fig:pfss}b renders the filling factor of mid-altitude loops as well as the location of high-altitude loops in the TPZ.
Table 2 contains a summary of the partitioning the TPZ.
First, let us consider the projection of mid- and high-altitude over the complete model area (labeled as Full Field in Table 2).
Our potential field contains 30 grid cells in the TPZ, and we identify an area as mid or high using the definition
\begin{multline*}
\mathrm{Mid\text{-}altitude~loop~region}=\\ \left(\sum_{z=1.5~\mathrm{Mm}}^{4.5 \mathrm{~Mm}} \mathrm{max}[h](z)<7 \mathrm{~Mm}\right)< 20
\end{multline*}
and
\begin{multline*}
\mathrm{High\text{-}altitude~loop~region}=\\ \left(\sum_{z=1.5~\mathrm{Mm}}^{4.5 \mathrm{~Mm}} \mathrm{max}[h](z)>7 \mathrm{~Mm}\right)< 20
\end{multline*}
While a majority of the full-field model area is covered by mid-altitude loops, if we only look at the areal regions overlying magnetic elements we find that high-altitude loops strongly dominate.
While this fits the canonical model of magnetic funnels, it does not provide us an explanation for why these regions would be bright.\\
\indent
We consider max[$h$]=7 Mm as the boundary between mid-altitude and high-altitude loops.
Our potential field allows us to map how this layer is connected to the photospheric boundary similar to \cite{schrijver_02}.
We find that only 14\% of our weak field source are associated with high-altitude loops.
This implies that we do not expect to find a dominant internetwork-coronal connection, although this quantity can vary with the total unsigned flux in the internetwork.
We also find that 41\% of the weak field regions do not have loops that reach the TPZ.
They close at heights below 1.5 Mm and will not effect TTP emission.\\
\indent Strong field flux concentrations (the extracted magnetic elements) vary in connectivity. 
By area, we find approximate equality between the regions connected to mid-altitude and high-altitude loops.
We find that the core of elements where high flux densities occur are more likely to connect to the high-altitude loops.
However, 30\% of the magnetic elements do not connect to any high-altitude loops.
This implies that close proximity network bipoles can exist without a coronal connection.\\
\indent The connectivity of our potential field and the Bifrost model differ significantly.
The mostly dipolar field of the Bifrost model does not contain the same division of scales as the real quiet sun.
To highlight the effect of mixed polarity magnetic fields in the potential field, we have extract four different regions of interest.
These regions, labeled B1-B4, were chosen because, based on the magnetic boundary data, all four could potentially be labeled as cell interior regions.
They are devoid of magnetic concentrations and the areas range between 400-1200 Mm$^2$,
similar to canonical supergranular cells.
B1 and B2 represent regions that contain very few short loops (average filling factor of 3\% and 30\% respectively).
B3 and  B4 represent regions that are dominated by short loops (average filling factor of 72\% and 95\% respectively).
While these two region groups are similar in scale and enclosed flux, they differ in the mixture of polarities.
In the potential field model, we observe a correlation between the presence of short loops and mixed polarity concentrations at the apparent cell vertices.
\begin{table*}
\centering
\begin{tabular}{c||c|c|c|c|c}
\multirow{2}{*}{Loop structure}& \multicolumn{2}{c|}{Projection in}&\multicolumn{2}{c|}{Connectivity \%$^\dagger$}& Mean $|B|$\\
& Full Field & Magnetic & Weak Field & Strong Field & at z=0 \\
\hline
Mid-altitude & 56\% & 25\% & 45\% &54\%& 52 G\\
High-altitude & 34\% &70\% &14\% &44\%& 70 G\\
\end{tabular}
\caption{Properties of the magnetic loops and the TPZ in the potential field model.~$^\dagger$Percentage of photospheric footpoints that are connected to loops of given height.}
\end{table*}
\section{Discussion}
In this study, we have sought to use both observations and models to probe the magnetic connectivity of the quiet sun.
We have initiated our work focussing on two widely agreed upon ideas that lacked detailed quantitative analysis.
\begin{itemize}
\item Observationally the magnetic network is linked to enhanced TTP and coronal emission, but there is no measured statistical relationship beyond flux-flux analysis.
\item The Dowdy model suggests that a mixture of loop types, high-altitude coronal loops and mid-altitude cool loops, populate the quiet sun, but observations have not been applied to quantify the partition of these components.
\end{itemize}
In Section 4 and 5, respectively, we have tried to expand on these ideas.
The evidence we have collected paints a complex and unintuitive picture.
Our analysis has not culminated in a smoking-gun piece of luminary evidence.
Rather we have found bits and fragments of evidence that allow us to develop a holistic model of quiet sun magnetic fields in the TPZ.\\
 \indent We posit the following constraints on that system based on the incorporation of evidence accumulated from analysis of both the observational data and the magnetic models:
 \begin{enumerate}[label=\Alph*)]
 \item The area of influence for network elements is limited. The network TTP emission contribution dominates no further than 6 Mm into the internetwork. The source of emission in this region is likely a combination of network-internetwork loops and spicules.
\item In the MHD model, mid-altitude loops are the dominant contributor to emission in the quiet sun, but the thermal conditions of individual loops vary drastically. These loops can connect to either weak or strong field regions.
\item In the MHD model, high-altitude loops are weak emitters.  Either the potential field model over-estimates the percentage of network flux that connects to the corona or high-altitude loops are more dynamic and variable in TTP emission than predicted by the MHD model.
\item Using the magnetic proximity statistics, we cannot identify a difference in the magnetic environment of network lanes versus cell interiors. We are not able to predict the locations of the dimmest TTP emission. One explanation is that connectivity of weak field regions is highly variable.
\end{enumerate}
We will discuss the evidence that developed each Constraint and its implications below.\\
\indent Constraint A addresses where the foot points of loops that thread the TPZ are rooted.
Our magnetic proximity statistics have provided evidence that strong flux concentrations are not solely responsible for quiet sun emission.
We found that there is not a continuous radial falloff when we look at relationship between radiance-$r^0$.
This is evidence that the internetwork generates its own TTP, which is consistent with the measurement of low-activity solar-analog stars \citep{rutten_91}.
If we use the magnetic distribution of our potential field model and stipulate that network-connected emission is exclusively responsible for TTP emission up to $r^0$=6.5 Mm, we estimate that the non-network contribution of emission would be over 25\%.
This is likely an underestimate as we expect that there is likely a gradual change in connectivity as we move radially outward from network elements.
We found basically no significant trend in radiance-$r^1$ for $r^1> 3$ Mm.
This suggests that magnetic loops that contain TTP across their entire length are unlikely to connect network elements more distantly separated.\\
\indent Constraints B and C are drawn primarily from our analysis of the MHD model.
We were able to use the simulation output to isolate the vertical swath of the atmosphere that is the source of TTP emission, characterize the loops that pass through that region, and compare loops in strong and weak field regions.
We find that 95\% of TTP emission comes from the two-scale-heights thick region 1.5 Mm$<z<$4.5 Mm (TPZ).
We find that the strongest emission within the TPZ occurs along loops where 2 Mm$<$max$[h]<$6 Mm.
These are the same loops recently described by \cite{hansteen_14}.
The high-altitude loops that the MHD model accurately captures, max$[h]<$14 Mm, have TTP emissivities 1-2 orders of magnitude lower than the mid-altitude loops.
We find that high-altitude loops emit less than 10\% of the total simulation TTP emission.
Thus any diagnostic of emission of TTP primarily provides information on UFS structure rather than coronal structure.
We have qualified Constraint B and C because these results wholly rely on the MHD simulation.
This is important because we know the MHD simulation differs from the Sun in at least two important ways.
First, we know that the amount of emission generated by the model underestimates the quiet sun emission by approximately an order of magnitude.
Second, we know that the model is unable to produce spicules with regularity (at least one per minute per flux system is an estimate from observations).
From a physical perspective, one can argue that these effects may be causally related.
Spicules are the manifestation of field aligned flows of chromospheric material.
While plasma is certainly transported vertically, it has not been determined how much mass stays suspended and for what duration of time.
The MHD model might lack the driving force that generates spicules, which in turn reduces the time-averaged density the the upper chromosphere and TPZ.
In the potential field model, most magnetic elements are covered by high-altitude loops.
If high-altitude loops are brighter than the MHD model predicts due to spicules than our loop models will better match the observations.
An alternative explanation can be conceived where the MHD model is accurately depicting the emission ratio between high- and mid-altitude loops, but the potential field model loop-partition statistics do not match the real Sun.
The solar chromosphere contains currents that change the projection of the TPZ relative to a potential field extrapolation.
If the cross-sectional expansion (as a function of height) of high-altitude loops is smaller than the potential field suggests, bright network elements could be explained by the dominance of mid-altitude loops in the TPZ.\\
\indent In our attempt to characterize quiet sun connectivity, we find that the unknowns of spicules create a stumbling block.
We do not know if spicules occur exclusively along high-altitude loops or if they also occur along more inclined mid-altitude loops.
In off-limb IRIS slitjaw data, spicule-producing regions are surrounded by diffuse emission.
It is likely that spicules and network-internetwork connected loops generate similar emission structures in long-exposure, on-disk observations.
Network-internetwork loops will be short and bright, based on the MHD results.
We know that there is likely a great deal of magnetic flux in the internetwork \citep{orozco_07}.
Constraint A indicates that heating in the distant internetwork is strong enough to produce TTP, and we expect that there will be more energy available nearer strong field regions.
Given the number of unknowns, we cannot differentiate network-internetwork loops and spicules relying solely on our analyzed data.\\
\indent While Constraints B and C focussed on what we learned about magnetic regions, Constraint D describes the TPZ environment above non-magnetic regions.
While the qualitative model of TTP that dates back to \cite{reeves_76} links network concentrations and bright TTP emission, the extension of that model to non-magnetic areas is less clear
Our data analysis shows that the magnetic proximity cannot tell us where the dimmest (cell interior) regions are located.
On average, cell interior regions are just as close to magnetic concentrations as network lanes but are 50\% dimmer.
This measurement implies that the TPZ connectivity in non-magnetic regions is linked to the magnetic flux distribution in a non-uniform way.
This observational result is backed up by our analysis of the potential field model.
When looking at the magnetic flux distribution of the photosphere, it seems routine to visually connect network flux concentrations and delineate non-magnetic cells 10-30 Mm across (regions comparable to supergranules).
However, we find that these regions vary wildly in terms of loops geometries in the TPZ.
This variation is tied to the flux balance over large scales (i.e. multiple cells) that cannot be quantified using our magnetic proximity statistic.
This effect may help explain the variability of the non-magnetic region radiance data.
Other aspects, such as the evolution and energization of magnetic fields through the Parker mechanism \citep{parker_72} could also play a role.\\
\indent In considering how to test the Dowdy quiet sun model, we ultimately decided that a large sample size of measurements was necessary.
We would need long exposures to compensate for the low count rates.
These requirements dictated that we could not address the role of dynamics in TTP.
We know that the chromosphere plasma is roiling on top of convection cells \citep{carlsson_97}.
We know that the the TTP emission is redshifted on average \citep{hardi_99}.
The TPZ is likely filled with plasma that is dynamically shifting in position and temperature.
Our analysis provides constraints on this process on average.
An ideal complementary study would be use rapid IRIS rasters to measure the time-dependent emission of Si IV surrounding network elements.
How far do spicules extend from the elements?
How frequent are they?
How does the TTP emission vary through the spicule's life cycle?
By coupling the dynamics results with our results on the time-independent large-scale structure, we can more accurately undercover the individual contribution of different loop structures and their thermal evolution.
Ultimately, these quantities will inform us on the connectivity of the quiet sun corona and provide the most rigorous constraint yet on the conduits of energy and mass.

\section*{}
We thank Karel Schrijver and Juan Martinez-Sykora for their helpful comments.
IRIS is a NASA small explorer mission developed and operated by LMSAL with mission operations executed at NASA Ames Research center and major contributions to downlink communications funded by ESA and the Norwegian Space Centre.
B.D.P. was supported by NASA grant NNX11AN98G and NASA contract NNG09FA40C (IRIS).
D.J.S was supported by NASA contract NNX12AN35A.






\end{document}